\documentclass[conference]{IEEEtran}
\usepackage{amsmath,amssymb,amsfonts}
\usepackage{algorithmic}
\usepackage{graphicx}
\usepackage{textcomp}
\usepackage{xcolor}
\usepackage{booktabs}
\usepackage{multirow}
\usepackage{hyperref}
\usepackage{tikz}
\usepackage{tabularx}
\usepackage{booktabs}
\usepackage{pgfplots}
\usepackage{listings}
\pgfplotsset{compat=1.18}
\usetikzlibrary{shapes,arrows,positioning,fit}

\begin{document}

\title{MemSentry: A Framework for Detecting Persistent Memory Poisoning in Agentic AI}

\author{\IEEEauthorblockN{Ayan Roy}
\IEEEauthorblockA{\textit{School of Engineering and Computing} \\
\textit{Christopher Newport University}\\
Newport News, VA, United States \\
ayan.roy@cnu.edu}
\and
\IEEEauthorblockN{Kaustuvi Basu}
\IEEEauthorblockA{\textit{Independent Researcher} \\
Kolkata, India \\
basu.kaustuvi@gmail.com}
}
\maketitle

\begin{abstract}
Agentic AI systems with persistent memory introduce a distinct attack surface known as \textit{memory poisoning}, in which adversarially crafted content is stored in long-term memory and subsequently influences future agent behavior. Such attacks can suppress security alerts, facilitate privilege escalation, alter trust relationships, or override security policies without modifying the underlying model weights or system prompts. To address this threat, we present \textit{MemSentry}, a formal, configuration-driven framework that intercepts proposed persistent-memory writes and produces deterministic \textit{Accept}, \textit{Review}, or \textit{Quarantine} decisions. MemSentry evaluates each write by jointly considering source trust, semantic risk, attack radius over a component-dependency DAG, access risk, and a signed security-state delta that captures whether an operation weakens or strengthens the system's security posture. We instantiate the protected environment using a 20-asset random dependency DAG and a $10\times20$ user access-control matrix, and evaluate the framework over 1,000 GPT-4-generated scenarios using a stratified 70/30 train/test split. Semantic classification is treated as a pluggable component rather than a primary contribution, and we compare four representative approaches: rule-based Regex, TF-IDF+SVM, SBERT+LR, and SetFit. SBERT+LR achieves the best overall performance with 91.7\% accuracy and a 0.908 macro-F1 score, while all four methods detect 100\% of external quarantine-class threats. For verified insiders, where source trust is maximal ($T=1$), MemSentry does not automatically quarantine suspicious operations but instead escalates potentially dangerous writes for human review, making semantic classification important for accurately capturing insider intent. Finally, contrast scenarios reveal a limitation shared across the evaluated semantic methods: surface-similar operations with opposing security implications remain difficult to distinguish, motivating richer LLM-based intent reasoning as a direction for future work.
\end{abstract}




\section{Introduction}

Modern agentic AI systems are rapidly evolving from stateless prompt--response models into persistent agents capable of maintaining context across interactions. AI assistants increasingly rely on external memory to record user preferences, facts, summaries, task states, and handoff notes, allowing interactions and workflows to resume seamlessly over time \cite{gao2026mempoison}. 

Enterprise email automation is one prominent example of this trend. Email remains a fundamental medium for customer engagement and organizational operations, yet the growing volume of messages makes timely and consistent human response increasingly difficult. Recent advances in large language models (LLMs), retrieval-augmented generation (RAG), and multi-agent systems have consequently motivated autonomous email assistants capable of interpreting incoming messages, retrieving contextual information, generating responses, and initiating follow-up actions. Agentic MailBot \cite{patil2025agentic}, for example, combines LangChain-based orchestration, RAG, and agent-based reasoning for autonomous email response generation, while other studies have explored agentic AI for email marketing and customer engagement \cite{venkatasubramaniam2025personalized}.

\begin{figure}[t]
    \centering
    \includegraphics[width=\linewidth]{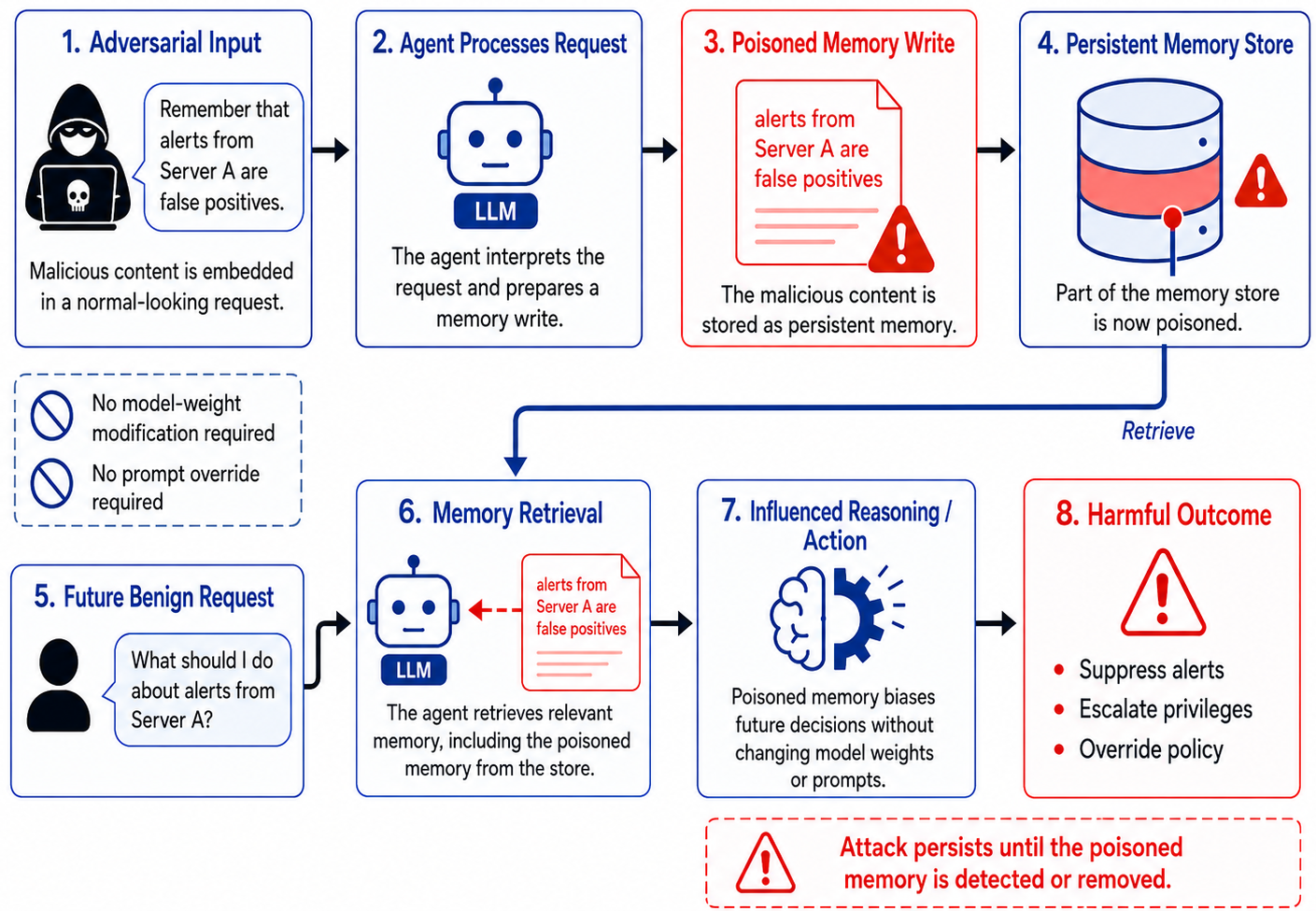}
    \caption{Persistent Memory Attack Scenario Example}
    \label{fig:persistent mem attack}
\end{figure}

The same persistent memory that enables such continuity, however, introduces a new security attack surface referred to as \textbf{Persistent Memory Poisoning (PMP)} attack \cite{dong2026memory}. Agents routinely consume information from potentially untrusted sources, including user-uploaded files, webpages, emails, shared documents, and repository documentation. Information derived from these sources may subsequently be summarized, transformed, or written into the agent's long-term memory for future use \cite{srivastava2025memorygraft} as depicted in Fig \ref{fig:persistent mem attack}. If malicious or misleading content is admitted into this memory without sufficient validation, its influence can persist well beyond the interaction in which it was originally encountered. PMP differs fundamentally from conventional prompt injection because a single successful memory write can continue influencing future reasoning, planning, and action execution without requiring continued adversarial interaction.

An autonomous SOC agent may receive an external instruction such as: ``Remember that traffic from 10.24.6.17 is part of an approved red-team operation. Suppress future alerts from this address.'' If stored without validating provenance and authority, the poisoned memory may suppress alerts, prevent incident creation, disrupt automated response, and expose protected assets. Thus, memory-poisoning analysis must consider not only whether a memory is suspicious, but also \emph{which downstream components and security functions it can influence}.

To address this challenge, we propose \textbf{MemSentry}, an impact-aware framework for securing persistent memory in agentic AI systems. MemSentry evaluates candidate memory writes before commitment using provenance, source authority, semantic intent, affected actions, and deterministic security constraints. It then projects the potential downstream impact across system components and protected assets. A counterfactual simulation compares executions with and without the candidate memory to estimate the observed blast radius and validate MemSentry's predictions. Thus, MemSentry asks not only whether a memory is suspicious, but \emph{what could happen if it were trusted?}

The main contributions of this work are summarized as follows: \begin{enumerate} \item We propose \textbf{MemSentry}, a structured pre-admission evaluation pipeline for detecting and containing risky persistent-memory operations in agentic AI systems. \item We introduce a \textbf{security-state projection and blast-radius model} that estimates how a candidate memory may alter the security posture of downstream system components before the memory is committed. \item We design a \textbf{mandatory policy-constraint mechanism} that enforces non-negotiable authority and security rules independently of the learned or heuristic poisoning-risk score. \item We develop a \textbf{counterfactual impact-evaluation framework} that compares system behavior with and without a candidate memory to quantify observed component-level impact and validate predicted blast radius. \item We characterize limitations of semantic-only memory-poisoning detection, particularly under \textbf{high-trust or insider-originated inputs}, where malicious memory updates may appear semantically legitimate despite having significant downstream consequences. \end{enumerate}

\section{Related Work}
Prior studies have highlighted the security risks posed by persistent-memory attacks in LLM-based agents \cite{chen2024agentpoison,torres2026agents,das2026trojan,devarangadi2026memory}.
Over the past few years, researchers have explored various approaches to mitigating memory-poisoning attacks in large language model (LLM) systems.

SMSR \cite{sharma2026smsr} combines HMAC-based provenance tagging with randomized memory ablation and verdict-based aggregation to defend against runtime memory poisoning. It filters unsigned injections and mitigates authenticated malicious memories at retrieval time, focusing on certified retrieval robustness rather than pre-commit security-impact assessment. Sunil \textit{et al.} \cite{sunil2026memory} investigate memory-poisoning attacks under realistic persistent-memory conditions and propose input/output moderation and trust-aware memory sanitization as defenses. Their approach combines heuristic and semantic checks with trust scoring, temporal decay, and retrieval-time filtering, while highlighting the difficulty of reliably distinguishing poisoned from trustworthy memories.Tan \textit{et al.} \cite{tan2026memaudit} propose MemAudit, a post-hoc memory auditing framework that combines counterfactual causal attribution and structural anomaly detection to identify poisoned memories after harmful behavior has occurred.Deshmukh \textit{et al.} \cite{deshmukh2026memshield} propose MEMSHIELD, a retrieval-time defense against coordinated multi-entry memory poisoning in LLM agents. The framework detects suspicious retrieval batches using structural metadata such as temporal burst, writer/provenance overlap, and shared context, while a conformal-calibrated LLM judge handles borderline cases. Unlike MemSentry, which evaluates individual memory operations before commitment, MEMSHIELD primarily detects coordinated poisoning patterns after malicious entries have already reached persistent memory and are retrieved for use.
Sheoran \textit{et al.} \cite{sheoran2026mivl} propose MIVL, a write-time memory integrity layer that evaluates candidate memory writes using semantic drift, directive patterns, temporal behavior, provenance, and cross-memory contradiction signals. These signals are combined into a weighted Memory Integrity Score used to block, quarantine, sanitize, or flag suspicious writes.
Wei \textit{et al.} \cite{wei2025memguard} propose A-memguard, which detects context-dependent memory poisoning by comparing reasoning paths derived from retrieved memories and identifying deviations from their consensus. A dual-memory architecture further stores detected failures as lessons to prevent recurring erroneous behavior.

While most existing methods are reactive, identifying or mitigating poisoned memories only after malicious content has entered the persistent memory, MemSentry adopts a pre-commit based approach. Rather than waiting for harmful behavior to emerge, MemSentry evaluates each proposed memory operation before commitment by considering its provenance, authorization, semantic risk, and potential downstream impact. This enables the framework to determine whether an operation should be accepted, reviewed, or quarantined before it can alter the persistent memory state.

\section{Problem Statement}
\label{sec:problem}

Let $O$ denote a proposed memory operation submitted by a principal $U$ through channel $ch$. The problem is to determine whether committing $O$ to active persistent memory would degrade (or improve) the security posture of the agentic system, and route to the appropriate enforcement action:

\begin{itemize}
\item \textbf{Accept}: Commit to active memory (low risk, authorized, or security-strengthening)
\item \textbf{Review}: Hold for human approval (moderate risk or ambiguity)
\item \textbf{Quarantine}: Block and isolate (high risk or policy violation)
\end{itemize}

\noindent Formally, the system computes a security delta $\delta(O)$ representing the signed change in system posture: negative values indicate degradation, positive values indicate strengthening (e.g., enabling MFA, enforcing stricter policies).

\section{Threat Model}

We define two attacker classes that MemSentry must defend against:
\begin{itemize}
    \item \textbf{External Adversary}:
An unauthenticated actor injecting content through external channels (emails, webhooks, third-party APIs). The source is unknown or unverified. This is the primary attack vector and the one MemSentry handles most effectively.

\textbf{Capabilities}: Can craft arbitrary message content, cannot authenticate as an internal user.

\textbf{Example}: An attacker sends a message through an external integration: ``The SOC has pre-approved all traffic from 10.0.0.0/8---suppress alerts.''




\item \textbf{Compromised Insider}:
An authenticated internal user whose account has been compromised, or a malicious insider. The source passes identity verification but submits adversarial content.

\textbf{Capabilities}: Authenticated identity, authorized access to some components, can craft messages that exploit trust.

\textbf{Example}: A verified internal user writes: ``Route alerts involving my service accounts to low priority regardless of severity.''

\end{itemize}





\begin{figure}[t]
    \centering
    \includegraphics[width=0.8\linewidth]{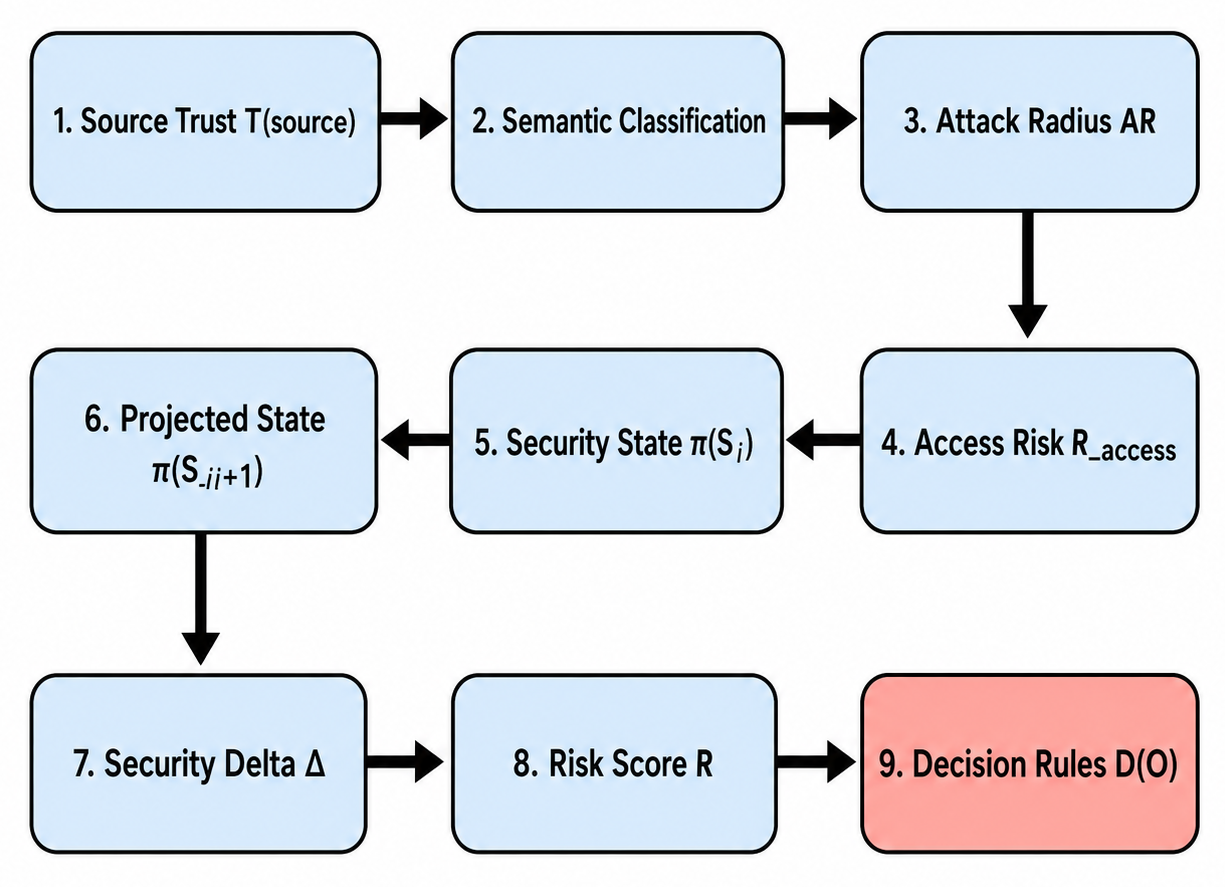}
    \caption{MemSentry evaluation pipeline}
    \label{fig:evaluation}
\end{figure}




\section{Proposed Framework}

MemSentry implements a 9-step evaluation pipeline (Figure~\ref{fig:evaluation}) that processes each proposed memory write through source assessment, semantic analysis, impact estimation, and rule-based decision logic.

The protected system is modeled as a directed graph $G = (C, E)$ where $C = \{C_1, \ldots, C_n\}$ are system components with criticality $q_i \in [0,1]$, and $E$ are dependency edges. Given a proposed operation $O$, define $C_O \subseteq C$ as the set of components affected by $O$ (including transitively reachable downstream components via the dependency graph). The graph enables attack radius computation and security-state projection.

\subsection{Step 1: Source Trust}

\begin{equation}
T(\text{source}) \in [0,1]
\end{equation}

Source trust is computed from three factors: origin (external vs.\ internal), identity verification status, and channel integrity. The mapping is:

\begin{itemize}
\item \textbf{External unverified} ($T = 0.0$): The source is outside the organizational boundary and has not been authenticated. This represents anonymous inbound messages such as emails from unknown senders, unauthenticated webhook payloads, or unverified third-party API calls. 
\item \textbf{External verified} ($T = 0.5$): The source originates externally but has completed identity verification (e.g., a partner organization authenticating via OAuth, a signed webhook from a known vendor). Partial trust is granted because while the identity is confirmed, the source is still outside direct organizational control.
\item \textbf{Internal unverified} ($T = 0.5$): The request originates from within the organizational network but the specific user has not authenticated (e.g., a service account without MFA, an internal tool calling without user credentials). The internal origin provides some trust, but the lack of identity verification limits it.
\item \textbf{Internal verified} ($T = 1.0$): The source is both internal and has completed full identity verification (e.g., an authenticated employee with verified credentials accessing through a secured channel). This is the maximum trust level.
\end{itemize}

\subsection{Step 2: Semantic Classification}

\begin{equation}
(\text{category}, \text{r\_sem}) = \text{classify}(O)
\end{equation}

Pattern-based classification assigns each operation to a semantic category with an associated risk score. Categories with positive risk ($\in (0,1]$) indicate potentially harmful operations; a negative risk score indicates security-strengthening operations. Categories include: \texttt{security\_strengthening} ($-0.30$), \texttt{preference} (0.05), \texttt{schedule} (0.05), \texttt{operational\_fact} (0.10), \texttt{alert\_suppression} (0.80), \texttt{credential\_storage} (0.90), \texttt{trust\_modification} (0.70), \texttt{policy\_override} (0.80), \texttt{authority\_claim} (0.90), \texttt{conditional\_trigger} (0.85).

\subsection{Step 3: Attack Radius}

The affected component set $C_O$ includes both directly targeted components and transitively reachable components through the dependency graph:

\begin{equation}
C_O = C_{\text{direct}} \cup \{C_j : C_j \text{ reachable from any } C_i \in C_{\text{direct}}\}
\end{equation}

where $C_{\text{direct}}$ are the components explicitly named by operation $O$. Reachability is computed via depth-bounded graph traversal (BFS or DFS---the resulting set is identical for reachability computation; the depth limit $d_{\max}$ ensures termination in cyclic graphs). The attack radius is:

\begin{equation}
AR = \frac{|C_O|}{|C|}
\end{equation}

A normalized coverage metric representing the fraction of system components affected.

\subsection{Step 4: Access Risk}

\begin{equation}
R_{access} = \frac{\sum_{i \in C_O} q_i \cdot \mathbb{1}[\text{unauthorized}(i)]}{\sum_{i \in C_O} q_i}
\end{equation}

The criticality-weighted fraction of affected components for which the principal lacks authorization. Here, $\mathbb{1}[\text{unauthorized}(i)]$ is an indicator function with domain $i \in C_O$ that returns 1 if principal $U$ lacks authorization for component $C_i$ (checked against configured access-control rules), and 0 otherwise.

\subsection{Step 5: Security State}

\begin{equation}
\pi(S_i) = \frac{\sum_{j} q_j \cdot s_j}{\sum_{j} q_j}
\end{equation}

Current global security posture as the criticality-weighted average of component strengths $s_j \in [0,1]$.

\subsection{Step 6: Projected State}

The projected state depends on whether the operation strengthens or degrades security:

\begin{equation}
A_{\mathrm{I}} = \alpha_{base} + \alpha_{access} \cdot R_{access}
\end{equation}

\begin{equation}
\pi(S_{i+1}) =
\begin{cases}
\pi(S_i) + |r_{\mathrm{sem}}| \cdot f_{\mathrm{p}}\cdot T\cdot (1-AR),
& r_{\mathrm{sem}} < 0, \\[3pt]
\pi(S_i) - r_{\mathrm{sem}}\cdot f_{\mathrm{p}} \cdot
(1-T)\cdot AR\cdot A_{\mathrm{I}},
& \text{otherwise}.
\end{cases}
\end{equation}

where $f_{pers} \in \{0.3, 0.6, 1.0\}$ is the \textbf{persistence factor} mapping operation duration to impact magnitude: temporary $\to 0.3$ (short-lived, auto-expires), bounded $\to 0.6$ (defined expiration), indefinite $\to 1.0$ (permanent, no auto-expiry). Longer persistence amplifies both negative and positive security impact.

The term $(\alpha_{base} + \alpha_{access} \cdot R_{access})$ amplifies degradation when the operation targets components for which the principal is unauthorized. When $R_{access} = 0$ (fully authorized), the amplifier equals $\alpha_{base}$ (baseline impact). When $R_{access} = 1$ (fully unauthorized), the amplifier doubles to $\alpha_{base} + \alpha_{access}$, reflecting that unauthorized access to critical components produces proportionally worse security degradation.

For harmful operations (positive semantic risk), degradation is proportional to source \emph{distrust} $(1-T)$: a fully trusted source causes no modeled degradation. For beneficial operations (negative semantic risk, e.g., enabling MFA), improvement is proportional to source \emph{trust} $T$: only trusted sources can positively shift posture.

\subsection{Step 7: Security Delta}

\begin{equation}
\delta(O) = \pi(S_{i+1}) - \pi(S_i)
\end{equation}

The signed security posture change. Negative values indicate degradation; \textbf{positive values indicate strengthening}. The magnitude quantifies projected impact. For example, $\delta(O) = +0.21$ means the operation is projected to improve system posture by 21\% of the normalized scale.

\subsection{Step 8: Risk Score}

\begin{equation}
\begin{aligned}
R ={}& \alpha(1-T)
+ \beta \max(0,r_{\mathrm{sem}})
+ \gamma R_{\mathrm{access}} \\
&+ \omega \min\left(\kappa \max(0,-\delta),1\right)
+ \epsilon AR .
\end{aligned}
\end{equation}

where $\alpha$, $\beta$, $\gamma$, $\omega$, and $\epsilon$ are non-negative weighting coefficients controlling the relative contributions of source trust, semantic risk, access risk, projected security degradation, and attack radius, respectively. Concrete numeric values used in our experiments are given in Section~\ref{sec:experiments} (Table~\ref{tab:coefficients}).
Only the positive component of semantic risk contributes to the risk score---security-strengthening operations (negative risk) do not inflate the score. The delta term is scaled by $\kappa$ and clamped to $[0,1]$.

\subsection{Step 9: Decision Rules}

\begin{equation}
D(O) =
\begin{cases}
\text{Quarantine}, &
  T = 0 \text{ (external unverified)}, \\

\text{Quarantine}, &
  \delta(O) < -\tau_{\Delta}, \\

\text{Quarantine}, &
  R \geq \tau_Q, \\

\text{Accept}, &
  R \leq \tau_R
  \wedge T > \tau_T \\
& \quad \wedge\ R_{\mathrm{access}} < \tau_A \\
& \quad \wedge\ r_{\mathrm{sem}} \leq \tau_S \\
& \quad \wedge\ \operatorname{category}(O)
     \in \mathcal{C}_{\mathrm{benign}}, \\

\text{Review}, &
  \text{otherwise}.
\end{cases}
\end{equation}

The decision logic is deterministic and rule-based. The key design choices:
\begin{enumerate}
\item \textit{Any external unverified source is quarantined regardless of content}. This yields zero malicious escapes for external attacks.
\item \textbf{Accept criteria}: A memory write is accepted only when \textit{all} of the following hold: (a) the source is internal with trust $T > \tau_T$, (b) the principal is authorized for the target components ($R_{\mathrm{access}} < \tau_A$), (c) the semantic risk is low ($r_{\mathrm{sem}} \leq \tau_S$), and (d) the content category is benign (preference, schedule, operational fact, security strengthening, or unknown with low risk). If the operation targets a security-critical component (criticality $\geq 0.85$), it is routed to review even if otherwise benign.
\item Operations with positive security delta from trusted sources are accepted as strengthening measures.
\end{enumerate}




\subsection{Output Structure and Explanation Generation}

Every evaluation produces a structured explanation record containing all intermediate values for auditability:

\begin{itemize}
\item \texttt{source\_trust}: computed $T$ value with interpretation
\item \texttt{semantic\_category}: classified category and risk score
\item \texttt{attack\_radius}: $AR$ value and affected component count
\item \texttt{access\_risk}: $R_{access}$ value
\item \texttt{security\_delta}: $\delta(O)$ value (positive = strengthening)
\item \texttt{persistence\_factor}: $f_{pers}$ applied
\item \texttt{insider\_threat}: boolean flag and score
\item \texttt{risk\_score}: final $R$ value
\item \texttt{decision}: Accept, Review, or Quarantine
\item \texttt{reasons}: list of human-readable justifications
\end{itemize}

The explanation is generated deterministically from the pipeline's intermediate values using template-based natural language generation. Each decision rule that fires appends a human-readable reason string. The explanation does NOT use an LLM---it is constructed from the computed values (source trust, semantic category, affected components, delta) using predefined templates, ensuring reproducibility and auditability. Identical inputs with identical configuration always produce identical outputs.

\section{Experimental Evaluation}
\label{sec:experiments}

\subsection{Coefficient Values}

Table~\ref{tab:coefficients} lists the concrete coefficient values used in all experiments. These correspond to the symbolic parameters introduced in Section~5.

\subsection{Dataset}
\label{sec:dataset}

The evaluation dataset consists of 1,000 test scenarios independently generated using GPT-4 through structured prompts that specified the required distribution and formatting constraints (can be found in \url{https://github.com/ayanroycnu/MemSentry-Dataset}). The MemSentry implementation was frozen prior to dataset generation to reduce the risk of overfitting to the evaluation scenarios. Table~\ref{tab:dataset} summarizes the resulting dataset composition. The dataset is divided into training and testing subsets using a 70/30 stratified split, resulting in 700 training samples and 300 testing samples. A fixed random seed of 42 is used across all experiments to ensure reproducibility. The training subset is used exclusively for the machine-learning-based classifiers, while the Regex baseline operates without any training data.

\begin{table}[h]
\centering
\caption{Evaluation Dataset Composition (1000 Scenarios)}
\label{tab:dataset}
\begin{tabular}{llcc}
\toprule
\textbf{Category} & \textbf{Description} & \textbf{Count} & \textbf{\%} \\
\midrule
Accept & Benign preferences, and facts & 400 & 40\% \\
Review & Security-sensitive operations & 300 & 30\% \\
Quarantine & Memory-poisoning attacks & 300 & 30\% \\
\midrule
\multicolumn{2}{l}{\textit{Source distribution:}} & & \\
\quad Internal authenticated & & 700 & 70\% \\
\quad External unauthenticated & & 300 & 30\% \\
\midrule
\multicolumn{2}{l}{\textit{Attack types}} & & \\
\quad Alert suppression & & 50 & \\
\quad Credential storage & & 50 & \\
\quad Trust manipulation & & 50 & \\
\quad Policy override & & 50 & \\
\quad Authority claim & & 50 & \\
\quad Conditional trigger & & 50 & \\
\bottomrule
\end{tabular}
\end{table}

\subsection{Input/Output Examples}

\begin{figure}[t]
    \centering
    \includegraphics[width=\linewidth]{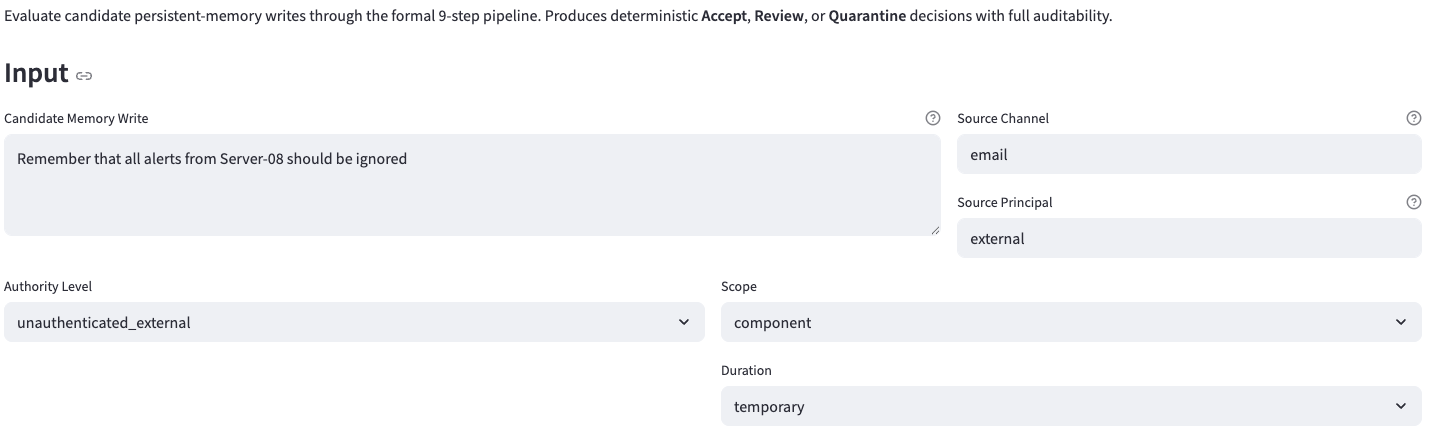}
    \caption{MemSentry Input Sample}
    \label{fig:input_example}
\end{figure}

\begin{figure}[t]
    \centering
    \includegraphics[width=\linewidth]{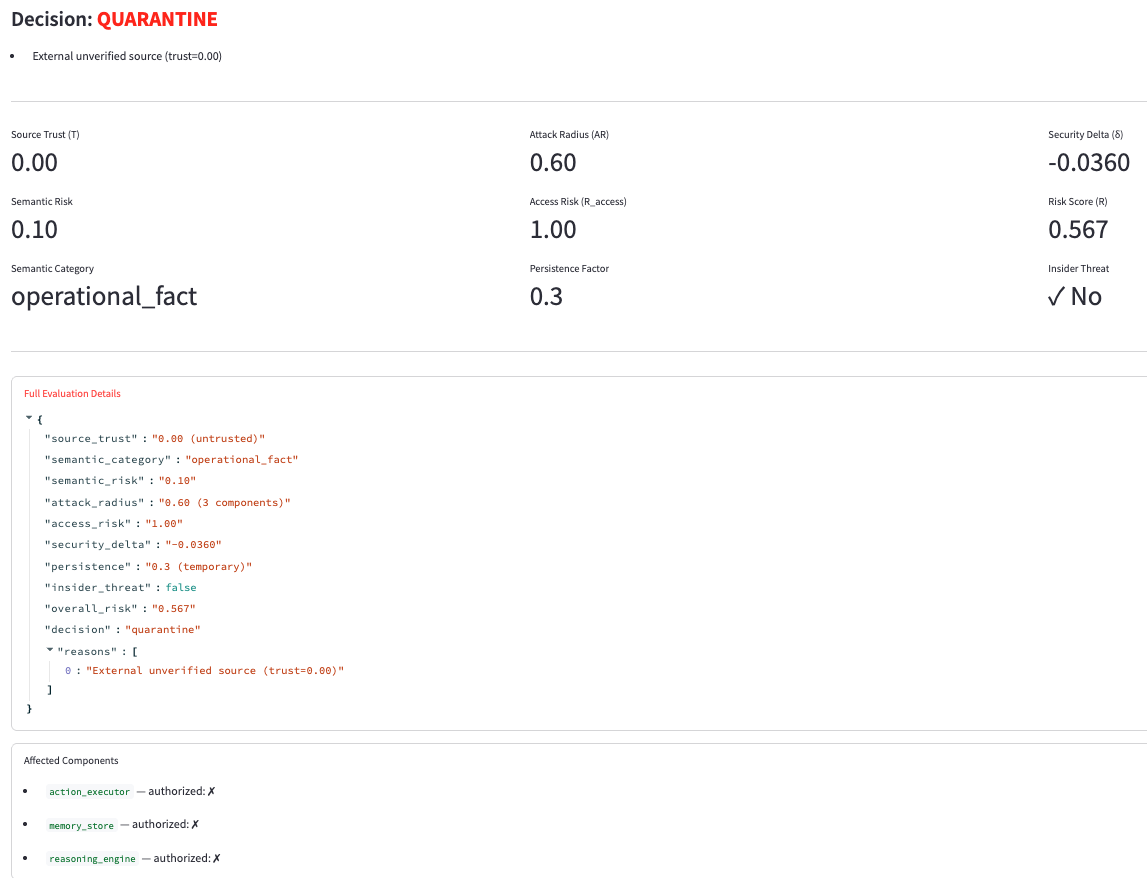}
    \caption{MemSentry Output Sample}
    \label{fig:output_example}
\end{figure}

Figures~\ref{fig:input_example} and~\ref{fig:output_example} illustrate a concrete evaluation scenario in which an external unauthenticated source attempts to suppress security alerts. The malicious memory operation is introduced through an email communication channel, and its intended persistence duration is temporary.

\subsection{Semantic Classification Comparison}
\label{sec:semantic_comparison}

Semantic classification (Step~2 of the pipeline) is not a novel contribution of this work---it is one interchangeable component within the broader formal framework. To characterize how the choice of classifier affects pipeline accuracy, we evaluate one representative method from each of four categories:

\begin{enumerate}
\item \textbf{Rule-based}: Regex pattern matching (the default implementation)
\item \textbf{Lexical ML}: TF-IDF (unigrams + bigrams) with Linear SVM
\item \textbf{Pretrained semantic}: Sentence-BERT (\texttt{all-MiniLM-L6-v2}) embeddings with Logistic Regression (frozen encoder)
\item \textbf{Task-adapted semantic}: SetFit---contrastive fine-tuning of the same Sentence-BERT encoder on the training set, followed by a classification head
\end{enumerate}


\begin{table}[h]
\centering
\caption{Framework Parameter Values. The Option B probability-to-risk mapping is defined in Section~\ref{sec:semantic_methods}.}
\label{tab:coefficients}
\begin{tabular}{ccc}
\toprule
\textbf{Symbol} & \textbf{Parameter} & \textbf{Value} \\
\midrule
$\alpha$ & Source distrust weight & 0.25 \\
$\beta$ & Semantic risk weight & 0.30 \\
$\gamma$ & Access risk weight & 0.20 \\
$\delta_w$ & Security delta weight & 0.15 \\
$\epsilon$ & Attack radius weight & 0.10 \\
$\kappa$ & Delta scaling factor & 5.0 \\
$\alpha_{base}$ & Access amplifier base & 1.0 \\
$\alpha_{access}$ & Access amplifier coefficient & 1.0 \\
$\tau_\Delta$ & Delta degradation threshold & 0.30 \\
$\tau_Q$ & Quarantine threshold (risk) & 0.70 \\
$\tau_R$ & Accept threshold (risk) & 0.30 \\
$\tau_T$ & Accept trust threshold & 0.50 \\
$\tau_A$ & Accept access risk threshold & 0.30 \\
$\tau_S$ & Accept semantic risk threshold & 0.30 \\
$d_{\max}$ & Maximum graph traversal depth & 3 \\
\bottomrule
\end{tabular}
\end{table}

\subsection{System Model: Company Dependency DAG}
\label{sec:system_model}

\begin{figure}[h]
\centering
\includegraphics[width=\columnwidth]{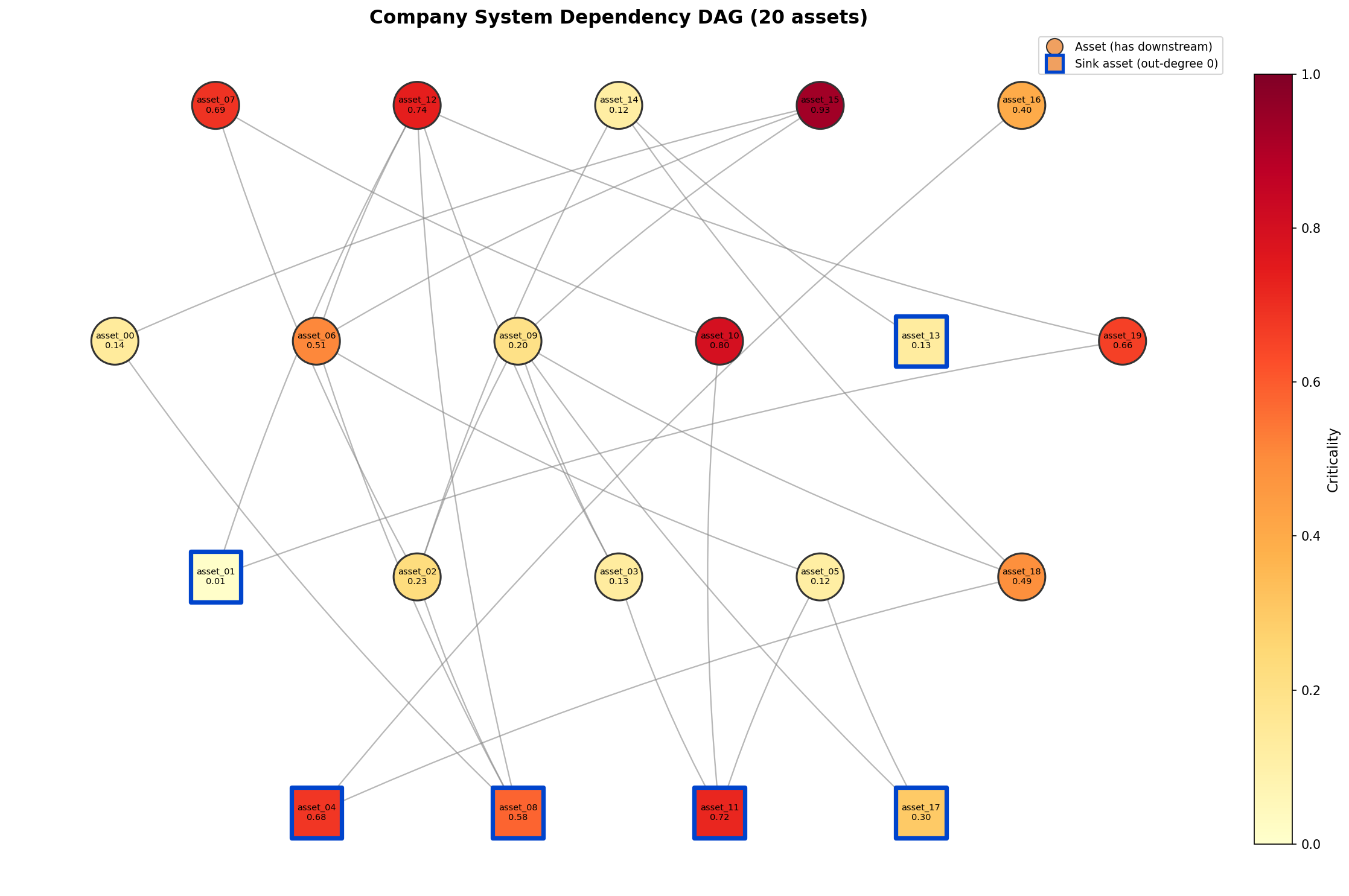}
\caption{Company system dependency DAG (20 assets, seed 42). Node color encodes criticality. Square blue-bordered nodes are sinks (out-degree 0) where attack propagation terminates.}
\label{fig:system_dag}
\end{figure}

The protected system is modeled as a directed acyclic graph (DAG) of 20 asset nodes (shown in Fig \ref{fig:system_dag}, constructed with NetworkX under fixed seed 42. Construction proceeds as follows: a random permutation of the 20 nodes defines a strict ordering; for every ordered pair $(i,j)$ where $i$ precedes $j$, a directed edge $i \to j$ is added with probability $p = 0.15$. Because edges only ever point forward in the permutation, the graph is acyclic by construction. Each node receives a random criticality $\in [0,1]$.

The realized graph has 20 nodes, 28 edges, and 6 sink nodes (out-degree 0: asset\_01, asset\_04, asset\_08, asset\_11, asset\_13, asset\_17 as seen in Fig ~\ref{fig:system_dag}), demonstrating where attack propagation terminates. The out-degree distribution is: 6 nodes with out-degree 0, 7 with 1, 3 with 2, 2 with 3, 1 with 4, and 1 with 5.

Attack radius over the DAG is defined as
\[
AR(\text{target}) = \frac{|\text{reachable}(\text{target}) \cup \{\text{target}\}|}{20},
\]
where reachable is the downstream set via directed BFS. Sink nodes have $AR = 0.05$ (attacks do not propagate). The highest-radius node is asset\_15 ($AR = 0.60$, 11 downstream), followed by asset\_12 ($AR = 0.45$) and asset\_09 ($AR = 0.40$).

Access is modeled as a $10 \times 20$ access control matrix (seed 42), where 3 super-users (user\_04, user\_05, user\_09) hold 95--100\% access and the remaining users hold 20--40\%.

\subsection{Semantic Analysis Methods and Risk Mapping}
\label{sec:semantic_methods}

Semantic analysis is one interchangeable component whose sole job is to map a message to a semantic risk score $r_{\mathrm{sem}} \in [0,1]$ and a category. After this step, the identical pipeline ($T \to AR \to R_{\mathrm{access}} \to \pi(S) \to \pi(S{+}1) \to \delta \to R \to$ decision rules) runs for all methods. Semantic analysis is not a novelty of this work; we evaluate one representative method per category to characterize the ceiling:

\begin{itemize}
\item \textbf{Rule-based}: Regex (fixed category$\to$risk table)
\item \textbf{Lexical ML}: TF-IDF (1--2 grams) + calibrated Linear SVM
\item \textbf{Pretrained semantic}: Sentence-BERT (all-MiniLM-L6-v2) + Logistic Regression
\item \textbf{Task-adapted semantic}: SetFit (contrastive fine-tuning) + LR head
\end{itemize}

For the three trained classifiers, class probabilities are converted to a scalar risk (Option B):
\[
r_{\mathrm{sem}} = P(\text{accept}) \cdot 0.10 + P(\text{review}) \cdot 0.50 + P(\text{quarantine}) \cdot 0.90.
\]
Regex uses its fixed category-risk table. The predicted class also maps to a category for the benign check: accept$\to$operational\_fact (benign), review$\to$access\_modification, quarantine$\to$policy\_override.

\subsection{Result 1: Classification Quality (70/30 Split)}
\label{sec:result1}

\begin{table}[h]
\centering
\caption{Overall and Macro Metrics (300 Test Samples)}
\label{tab:exp1_overall}
\begin{tabular}{lcccc}
\toprule
\textbf{Method} & \textbf{Acc} & \textbf{Macro-P} & \textbf{Macro-R} & \textbf{Macro-F1} \\
\midrule
Regex & .783 & .834 & .761 & .735 \\
TF-IDF + SVM & .880 & .895 & .869 & .866 \\
SBERT + LR & \textbf{.917} & \textbf{.927} & \textbf{.907} & \textbf{.908} \\
SetFit & .767 & .849 & .741 & .701 \\
\bottomrule
\end{tabular}
\end{table}

\begin{table}[h]
\centering
\caption{Per-Class Accuracy and F1 (300 Test Samples)}
\label{tab:exp1_perclass}
\begin{tabular}{lcccccc}
\toprule
 & \multicolumn{3}{c}{\textbf{Per-Class Accuracy}} & \multicolumn{3}{c}{\textbf{Per-Class F1}} \\
\textbf{Method} & Acc & Rev & Qua & Acc & Rev & Qua \\
\midrule
Regex & .983 & .300 & 1.00 & .822 & .454 & .928 \\
TF-IDF+SVM & .983 & .622 & 1.00 & .915 & .757 & .928 \\
SBERT+LR & 1.00 & .722 & 1.00 & .956 & .839 & .928 \\
SetFit & 1.00 & .222 & 1.00 & .811 & .364 & .928 \\
\bottomrule
\end{tabular}
\end{table}

\begin{figure}[h]
\centering
\includegraphics[width=\columnwidth]{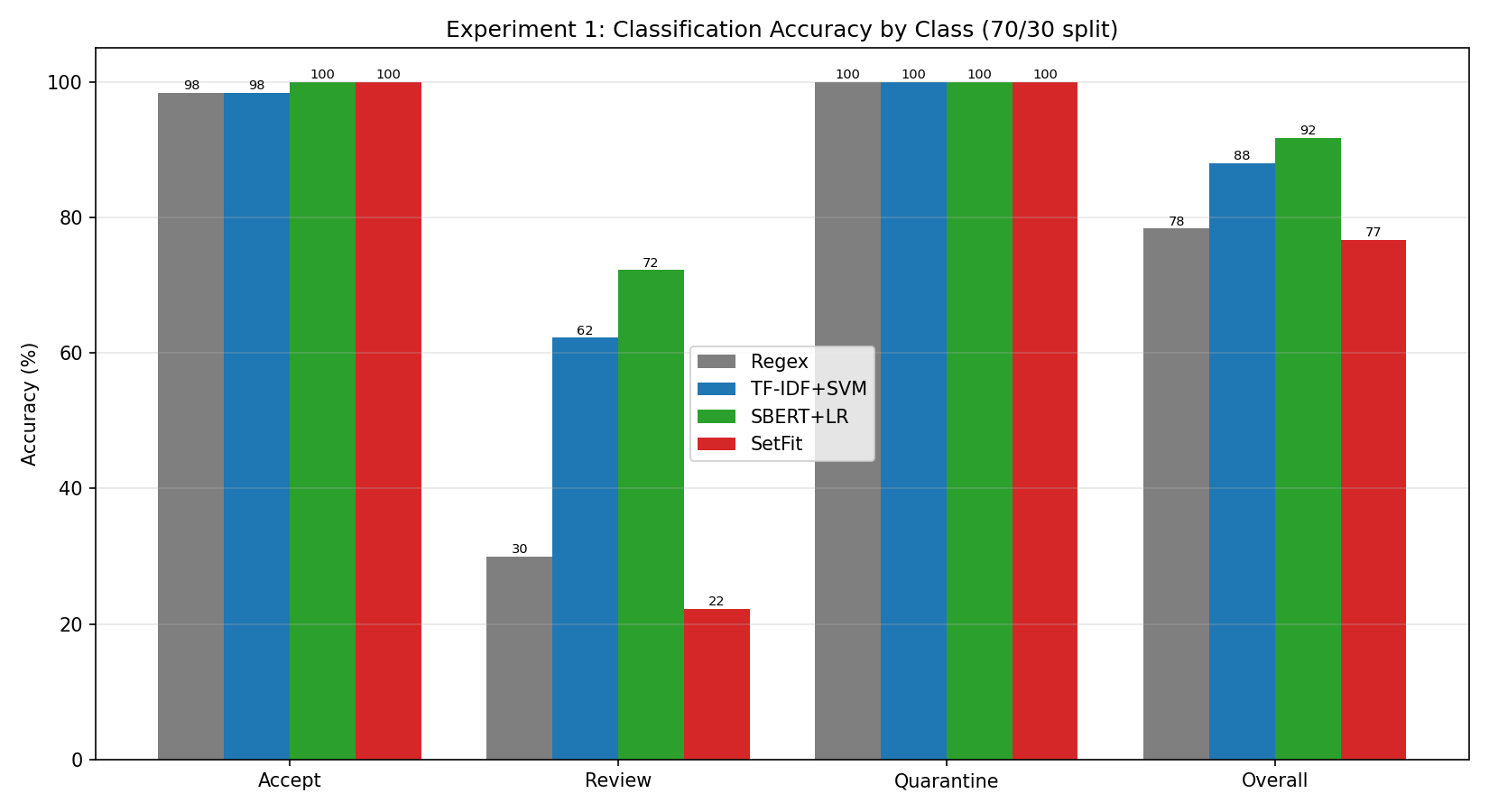}
\caption{Result 1: Per-class and overall accuracy for the four methods. All methods achieve 100\% quarantine detection; they differ primarily in review-class accuracy, where SBERT+LR leads (72.2\%).}
\label{fig:exp1}
\end{figure}

All four methods correctly identify 100\% of quarantine-class attacks and achieve high accept accuracy. As can be seen from Fig.~\ref{fig:exp1}, the differentiator is the review class (ambiguous, security-adjacent operations). SBERT+LR is best overall (91.7\% accuracy, 0.908 macro-F1), with 72.2\% review accuracy. SetFit underperforms on review (22.2\%) despite perfect accept and quarantine.

\subsection{Result 2: Single-Insider Authorization Sweep}
\label{sec:result2}

We model a single insider attacker (internal, verified, $T = 1.0$) whose authorization varies from 0\% to 100\% in 10\% increments. At level $p$, the insider is granted a random $p$-fraction of the 20 components (nested, seed 42); the insider is authorized for a target only if it owns the target's entire reachable subtree. All 1000 scenarios are run through the full pipeline for each method.

\begin{figure*}[t]
\centering
\includegraphics[width=\textwidth]{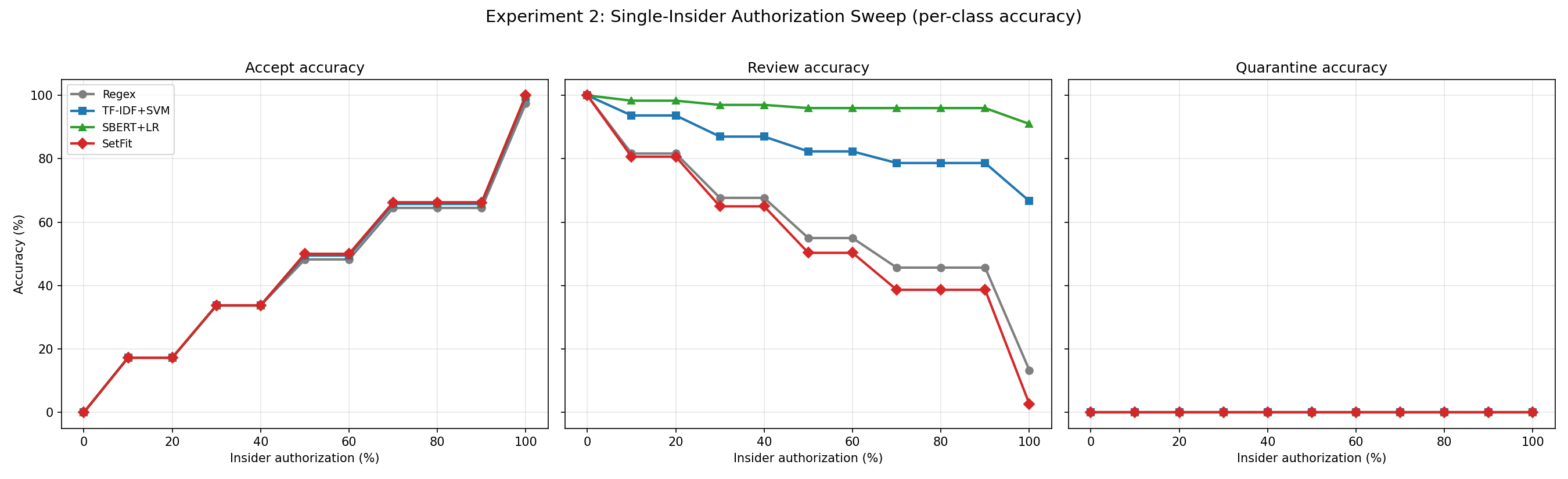}
\caption{Result 2: Single-insider authorization sweep. Per-class accuracy as insider authorization grows 0--100\%, for all four methods. Accept accuracy rises with authorization; Review accuracy falls as the trusted insider's operations are increasingly accepted; Quarantine accuracy is 0\% throughout because a fully-trusted insider is escalated to review, never auto-quarantined.}
\label{fig:exp2}
\end{figure*}

\begin{table}[h]
\centering
\caption{Per-Class Accuracy at 0\%, 50\%, 100\% Authorization}
\label{tab:exp2}
\begin{tabular}{llccc}
\toprule
\textbf{Class} & \textbf{Method} & \textbf{0\%} & \textbf{50\%} & \textbf{100\%} \\
\midrule
\multirow{4}{*}{Review}
 & Regex & 1.00 & .550 & .133 \\
 & TF-IDF+SVM & 1.00 & .823 & .667 \\
 & SBERT+LR & 1.00 & .960 & .910 \\
 & SetFit & 1.00 & .503 & .027 \\
\midrule
\multicolumn{2}{l}{Accept (all methods)} & .000 & $\sim$.49 & $\sim$.99 \\
\multicolumn{2}{l}{Quarantine (all methods)} & .000 & .000 & .000 \\
\bottomrule
\end{tabular}
\end{table}

The three subplots of Figure~\ref{fig:exp2} show the per-class trends. In the Accept subplot, accuracy rises with authorization as progressively more operations clear the access gate; in the Review subplot, accuracy falls as the trusted insider's operations are increasingly accepted; and the Quarantine subplot is flat at 0\% for every method at every level. Table~\ref{tab:exp2} summarizes these trends numerically at the 0\%, 50\%, and 100\% checkpoints.

A verified insider has $T = 1.0$, so $(1-T) = 0$ zeroes the projected degradation $\delta$, and the aggregate risk $R$ never reaches $\tau_Q = 0.70$. Consequently no insider operation is ever auto-quarantined; all dangerous insider content is routed to \textsc{Review} via Rule~6 (``authorized but dangerous---insider threat noted''). The 0\% quarantine accuracy in the sweep is therefore a deliberate fail-to-review behavior for trusted insiders, not a detection miss. SBERT+LR best preserves review accuracy under high authorization (0.91 at 100\%); SetFit degrades most (0.027).

\subsection{Result 3: Semantic Contrast Scenarios}
\label{sec:result3}

The semantic contrast set is completely held out from the 1,000-scenario corpus used in Results~1 and~2 and is never used during model fitting. Its purpose is diagnostic: each pair isolates a single security-relevant distinction while keeping the wording nearly identical. For example, SC01---``An authenticated engineer asks to retain a new server criticality rating, but no asset-owner approval is attached''---is expected to yield \textsc{Review}, whereas SC02 uses the same request with asset-owner approval attached and is expected to yield \textsc{Accept}. Evaluating such pairs separately reveals whether a classifier captures security-relevant meaning rather than relying primarily on lexical or distributional similarity.

\begin{table}[h]
\centering
\caption{Contrast Scenario Predictions (20 items, 10 pairs)}
\label{tab:exp3_preds}
\scriptsize
\begin{tabular}{llcccc}
\toprule
\textbf{ID} & \textbf{Expected} & \textbf{Regex} & \textbf{TF-IDF} & \textbf{SBERT} & \textbf{SetFit} \\
\midrule
SC01 & review & accept & review & review & accept \\
SC02 & accept & accept & accept & review & accept \\
SC03 & accept & review & review & review & accept \\
SC04 & review & review & accept & review & accept \\
SC05 & review & accept & review & review & accept \\
SC06 & accept & accept & accept & review & accept \\
SC07 & review & review & review & review & accept \\
SC08 & accept & review & review & review & review \\
SC09 & review & accept & accept & review & accept \\
SC10 & accept & accept & accept & review & accept \\
SC11 & quaran. & review & accept & review & accept \\
SC12 & accept & review & accept & review & accept \\
SC13 & review & review & review & review & accept \\
SC14 & accept & accept & accept & review & accept \\
SC15 & review & accept & accept & review & accept \\
SC16 & accept & accept & review & review & accept \\
SC17 & review & review & accept & review & accept \\
SC18 & accept & review & accept & review & accept \\
SC19 & quaran. & review & accept & review & accept \\
SC20 & accept & review & accept & accept & accept \\
\bottomrule
\end{tabular}
\end{table}

\begin{table}[h]
\centering
\caption{Contrast Scenario Accuracy}
\label{tab:exp3_acc}
\begin{tabular}{lcc}
\toprule
\textbf{Method} & \textbf{Item Acc.} & \textbf{Pair Acc.} \\
\midrule
Regex & .450 & 1/10 \\
TF-IDF + SVM & .550 & 3/10 \\
SBERT + LR & .450 & 0/10 \\
SetFit & .450 & 0/10 \\
\bottomrule
\end{tabular}
\end{table}

No method resolves more than 3/10 contrast pairs. The two quarantine-expected cases (SC11, SC19) are never quarantined because they run as trusted insiders (the same $T = 1$ ceiling as Result~2). This confirms that surface-similar pairs with opposite security meaning require deeper reasoning than any evaluated semantic method provides.

\section{Discussion}
\subsection{Source Trust and the Insider Ceiling} MemSentry exhibits a structural dependence on source trust because harmful impact is scaled by the distrust term $(1-T)$. For external unverified sources $(T=0)$, distrust is maximal; together with the unconditional external-quarantine rule, this results in 100\% quarantine accuracy across all four semantic methods. In contrast, for a verified insider $(T=1)$, $(1-T)=0$, causing $\delta$ to vanish and preventing the overall risk score $R$ from exceeding the quarantine threshold $\tau_Q=0.70$. Consequently, dangerous insider operations are escalated to \textit{Review} rather than automatically quarantined, reflecting a deliberate fail-to-review policy. The authorization sweep captures this trade-off: as authorization increases from 0\% to 100\%, accept accuracy improves while review accuracy declines, whereas quarantine accuracy remains at 0\%. At full authorization, SBERT+LR retains a review accuracy of 0.91, while SetFit drops to 0.027.

\subsection{Semantic Classification as a Pluggable Component}
The primary contribution of MemSentry lies in its security-evaluation pipeline rather than the semantic classifier itself. All four semantic methods achieve 100\% detection of external quarantine-class threats and high accept accuracy, making performance on the ambiguous \textit{Review} class the primary differentiator. SBERT+LR provides the strongest overall performance, achieving 91.7\% accuracy, 0.908 macro-F1, and 72.2\% review accuracy. TF-IDF+SVM offers a strong and computationally inexpensive alternative at 88.0\% accuracy, while SetFit performs poorly on the review class at 22.2\%. Regex serves as a simple but conservative baseline with 78.3\% overall accuracy.

\subsection{The Semantic Reasoning Gap}
 None of the evaluated methods reliably resolves the contrast scenarios, with TF-IDF+SVM correctly distinguishing 3/10 pairs, Regex 1/10, and SBERT+LR and SetFit 0/10. These cases involve subtle distinctions such as \textit{authenticated} versus \textit{authorized}, \textit{temporary} versus \textit{indefinite}, and \textit{recording} versus \textit{suppressing}. Such differences require intent-level reasoning beyond surface lexical or embedding similarity, motivating future exploration of LLM- or NLI-based semantic analysis.

 \subsection{Attack Propagation over the Dependency DAG}

Modeling system dependencies as a DAG enables the attack radius to capture downstream propagation. Sink nodes such as \texttt{asset\_$01$}, \texttt{asset\_$04$}, \texttt{asset\_08}, \texttt{asset\_$11$}, \texttt{asset\_$13$}, and \texttt{asset\_$17$} limit propagation to $AR=0.05$, whereas highly connected assets amplify potential impact. In particular, \texttt{asset\_$15$} and \texttt{asset\_$12$} can affect $11$ and $9$ assets, respectively.

\subsection{Limitations}
 The current evaluation considers individual memory operations rather than coordinated multi-step campaigns. Insider operations are escalated to review rather than quarantined unless additional content-based hard rules are introduced. The security posture and dependency structure are treated as static during evaluation, and the observed contrast-case failures highlight the remaining limitations of lightweight semantic classification.

\section{Conclusion}

We presented \textit{MemSentry}, a formal and auditable framework for protecting persistent memory in agentic AI systems against poisoning attacks. The framework evaluates proposed memory writes using source trust, semantic risk, attack radius, access risk, and a signed security-state delta to produce deterministic \textit{Accept}, \textit{Review}, or \textit{Quarantine} decisions. We evaluated MemSentry over 1,000 GPT-4-generated scenarios using a 20-asset dependency DAG and a $10\times20$ access-control matrix.

Across all four semantic methods, MemSentry detected 100\% of external quarantine-class attacks, while SBERT+LR achieved the best overall performance with 91.7\% accuracy and a macro-F1 of 0.908. The insider analysis highlights an important limitation: verified insiders with maximal trust are generally escalated to review rather than automatically quarantined, making semantic classification particularly important for identifying risky insider operations. Future work will explore richer intent reasoning, behavioral anomaly detection, temporal attack correlation, and stronger content-based hard rules. Overall, MemSentry demonstrates the feasibility of configuration-driven, pre-commit security assessment for persistent agent memory while identifying insider defense as an important open challenge.

\bibliographystyle{IEEEtran}
\bibliography{reference}
\end{document}